\documentclass[preprint]{revtex4}
\usepackage{epsfig}
\usepackage{graphicx}%
\usepackage{dcolumn}
\usepackage{amsmath}

\makeatletter
\def\btt#1{\texttt{\@backslashchar#1}}%
\DeclareRobustCommand\bblash{\btt{\@backslashchar}}%
\makeatother

\begin{document}

\title{Phantom with Born-Infeld type Lagrangian}

\author{Jian-gang Hao}
\author{Xin-zhou Li}\email{kychz@shtu.edu.cn}
\affiliation{SUCA, Shanghai United Center for Astrophysics,
Shanghai Normal University, 100 Guilin Road, Shanghai 200234,China
}%

\date{\today}

\begin{abstract}
ABSTRACT: Recent analysis of the observation data indicates that
the equation of state of the dark energy might be smaller than
$-1$, which leads to the introduction of phantom models featured
by its negative kinetic energy to account for the regime of
equation of state $w<-1$. In this paper, we generalize the idea to
the Born-Infeld type Lagrangian with negative kinetic energy term
and give the condition for the potential, under which the late
time attractor solution exists and also analyze a viable
cosmological model in such a scheme.
\end{abstract}

 \maketitle

\vspace{0.4cm} \noindent \textbf{1. Introduction} \vspace{0.4cm}

More and more astronomical observations converge on that about two
thirds of the energy density in our universe is resulted from dark
energy that has negative pressure and can drive the accelerating
expansion of the universe\cite{newobservation}. Many candidates
for dark energy have been proposed so far to fit the current
observations. Among these models, the most important ones are
cosmological constant and a time varying scalar field evolving in
a specific potential, referred to as
"quintessence"\cite{steinhardt}. The major difference among these
models are that they predict different equation of state of the
dark energy and thus different cosmology. Especially, for these
models, the equations of state are confined within the range of
$-1<w<-1/3$, which can drive the accelerating expansion of the
universe. However, some analysis to the observation data hold that
the range of the equation of state may not always be greater than
$-1$, in fact, they can lie in the range
$-1.38<w<-0.82$\cite{melchiorri}. Especially, the new results from
SN-Ia alone are suggesting $w<-1$ at 1 $\sigma$\cite{tonry}. It is
obvious that the equation of state of conventional quintessence
models that based on a scalar field with positive kinetic energy
can not evolve to the the regime of $w<-1$, and therefore, some
authors\cite{caldwell1,sahni,parker,chiba,boisseau,schulz,faraoni,maor,onemli,
torres,carroll,frampton,hao,caldwell2,gibbons,feinstein}
investigated phantom field models that possess negative kinetic
energy and can realize $w<-1$ in their evolution. It is true that
the field theory with negative kinetic energy poses a challenge to
the widely accepted energy condition and leads to rapid vacuum
decay\cite{carroll}, but it is still very interesting to study
these models in the sense that it is phenomenologically
interesting.

On the other hand, the role of rolling tachyon in string theory in
cosmology has been widely studied\cite{tachyon}. It is shown that
the tachyon can be described by a Born-Infeld (B-I) type
Lagrangian with a specific potential resulted from string theory.
However, it is known that the tachyon field is not suitable for
driving the late time accelerating expansion of the universe
because the dynamical evolution does not admit a late time
attractor solution\cite{li}. In this paper, we combine the two
ideas together and introduce the negative kinetic energy term to
the B-I type Lagrangian and investigate its cosmological
evolution. We firstly study the B-I type phantom model in an
arbitrary potential and give the condition for the potential to
admit a late time attractor solution that corresponds to the
equation of state $w=-1$ and then work in a specific viable model.
We also demonstrate that current universe is not a stable stage in
such a model while is in its way to the stable stage, at which the
universe is dominated by the vacuum energy like dark energy with a
equation of state of $-1$.

\vspace{0.4cm} \noindent\textbf{2. The Phantom Model With B-I
Lagrangian}
 \vspace{0.4cm}

Although there exist a number of impressive calculations of the
phantom field, yet the precise Lagrangian is not completely known
at present. Among the many models that have been proposed to
account for the dark energy in the universe, k-essence is the most
general form\cite{gibbons}, in which the lagrangian is expressed
as $L=L(\phi, y)$, where
$y=\frac{1}{2}g^{\mu\nu}\partial_{\mu}\phi\partial_{\nu}\phi$.
Sen's tachyon theory corresponds to the choice of Lagrangian as
$L=-V(\phi) \sqrt{1+2y}$. Gibbons\cite{gibbons1} has argued that
there are many objections to a naive inflationary model based on
the tachyon, but there remains the possibility that the tachyon
was important in a possible pre-inflationary "open-string era"
preceding our present "closed-string ear". In this paper, we
consider the case that the kinetic energy term is negative
\begin{equation}\label{lagoftp} L=-V(\phi)
\sqrt{1-2y}
\end{equation}

\noindent in the spatially flat Robertson-Walker metric,
\begin{equation}
ds^{2}=-dt^{2}+a^{2}(t)(dx^{2}+dy^{2}+dz^{2})
\end{equation}

\noindent Then for the spatially homogeneous scalar field, we have
the following Lagrangian

\begin{equation}
L=-V(\phi)\sqrt{1+\dot{\phi}^2}
\end{equation}

\noindent where $V(\phi)$ is the potential of the model. When we
consider the phantom field $\phi$ dominant era, the Einstein
equations for the evolution of the background
metric,$G_{\mu\nu}=\kappa T_{\mu\nu}$ can be written as:

\begin{equation}\label{H}
H^2=(\frac{\dot{a}}{a})^2=\frac{\kappa}{3}\rho_{\phi}
\end{equation}
\noindent and
\begin{equation}\label{dH}
\frac{\ddot{a}}{a}=-\frac{\kappa}{3}(\frac{1}{2}\rho_{\phi} +
\frac{3}{2}p_{\phi})
\end{equation}

\noindent For a spatially homogenous phantom field $\phi$, we have
the equation of motion
\begin{equation}\label{T}
\ddot{\phi}
+3H\dot{\phi}(1+\dot{\phi}^2)-\frac{V^{'}(\phi)}{V(\phi)}(1+\dot{\phi}^2)=0
\end{equation}

\noindent where the over dot represents the differentiation with
respect to $t$ and the prime denotes the differentiation with
respect to $\phi$. Eq.(\ref{T}) is also equivalent to the entropy
conservation equation. The constant $\kappa=8\pi G$ where $G$ is
Newtonian gravitation constant. The density $\rho_{\phi}$ and the
pressure $p_{\phi}$ are defined as following:

\begin{equation}
\rho_{\phi}=\frac{V(\phi)}{\sqrt{1+\dot{\phi}^2}}
\end{equation}

\begin{equation}
p_{\phi}=-V(\phi)\sqrt{1+\dot{\phi}^2}
\end{equation}

\noindent the equation of state is

\begin{equation}
w_{\phi}=\frac{p_{\phi}}{\rho_{\phi}}=- \dot{\phi}^2-1
\end{equation}

\noindent It is clear that the equation of state $w$ will be less
than $-1$ unless the kinetic energy term $\dot{\phi}^2=0$.

\vspace{0.4cm} \noindent \textbf{3. Dynamical Evolution of Phantom
Field} \vspace{0.4cm}

In this section, we investigate the global structure of the
dynamical system via phase plane analysis and compute the
cosmological evolution by numerical analysis. Firstly, we consider
the evolution when the phantom field becomes dominant and thus
neglect the non-relativistic and relativistic components (matter
and radiation) in the universe. Then, from Eq.(\ref{H}) and
(\ref{T}), we have
\begin{equation}\label{maseq}
\ddot{\phi}
+\sqrt{3\kappa}\dot{\phi}\left[\frac{V(\phi)}{\sqrt{1+\dot{\phi}^2}}\right]^{1/2}
(1+\dot{\phi}^2)-\frac{V^{'}(\phi)}{V(\phi)}(1+\dot{\phi}^2)=0
\end{equation}

To gain more insight into the above equation while not lose
generality, we here do not specify the potential. Introducing the
new variables
\begin{eqnarray}\label{newvar}
X=\phi\\\nonumber Y=\dot{\phi}
\end{eqnarray}

\noindent then Eq.(\ref{maseq}) becomes

\begin{eqnarray}\label{auto}
\frac{d X}{dt}&&=Y\\\nonumber \frac{d Y}{dt}&&=(1+Y^2)
\frac{V'}{V}-\sqrt{3V\kappa} (1+Y^2)^{3/4}Y
\end{eqnarray}

\noindent Linearize the above equation around its critical point
$(X_c, 0)$, one obtain that
\begin{eqnarray}\label{linear}
\frac{dX}{dt}&&= Y\\\nonumber \frac{dY}{dt}&&=
\frac{V''(X_c)}{V(X_c)}X-\sqrt{3V(X_c)\kappa}Y
\end{eqnarray}
\noindent where the critical value of $X$ is determined by
$V'(X_c)=0$. The types of the critical point is determined by the
eigenequation of the system
\begin{equation}\label{eigeneq}
\lambda^2+\alpha\lambda+\beta=0
\end{equation}
\noindent where $\alpha=\sqrt{3V(X_c)}\kappa$ and
$\beta=-\frac{V''(X_c)}{V(X_c)}$. The two eigenvalues are
\begin{equation}\label{eigenvalue1}
 \lambda_1=\frac{1}{2}\left(-\sqrt{3V(X_c)\kappa}-
 \sqrt{3V(X_c)\kappa+4\frac{V''(X_c)}{V(X_c)}}\right)
\end{equation}

\begin{equation}\label{eigenvalue2}
 \lambda_1=\frac{1}{2}\left(-\sqrt{3V(X_c)\kappa}+
 \sqrt{3V(X_c)\kappa+4\frac{V''(X_c)}{V(X_c)}}\right)
\end{equation}

\noindent For positive potentials, If $V''(X_c)<0$, then the
critical point $(X_c, 0)$ is a stable node, which implies that the
dynamical system admits attractor solutions.

In the following, we will show this with a specific model, to do
which, we must specify the potential. We choose the widely studied
potential for tachyon as\cite{kutasov}
\begin{equation}\label{poten}
V(\phi)=V_0(1+\frac{\phi}{\phi_0})\exp(-\frac{\phi}{\phi_0})
\end{equation}

\noindent It is not difficult to find that the critical $X_c=0$
and $\frac{V''(X_c)}{V(X_c)}=-\frac{1}{\phi_0^2}$ in such a
potential. Therefore, this model has an attractor solution which
corresponds to $\phi=0$, $\dot{\phi}=0$ and thus $w=-1$. Before
the field evolves to its attractor regime, the equation of state
$w<-1$.

Next, we solve this model via numerical approach. To do this, we
firstly re-scale the quantities as $\phi=x\phi_0$ and $t=s\phi_0$.
Then the Eq.(\ref{auto}) becomes dimensionless

\begin{eqnarray}\label{dimlessauto}
\frac{d x}{ds}&&=y\\\nonumber \frac{d
y}{ds}&&=-\frac{x(1+y^2)}{1+x}-\eta
y(1+y^2)^{\frac{3}{4}}(1+x)^{\frac{1}{2}}\exp(-\frac{x}{2})
\end{eqnarray}

\noindent where $\eta=\sqrt{3V_0\kappa}\phi_0$ is a dimensionless
parameter. The numerical results are plotted in Fig.1, Fig.2,
Fig.3 and the parameter $\eta=2.1$.

\begin{figure}
\epsfig{file=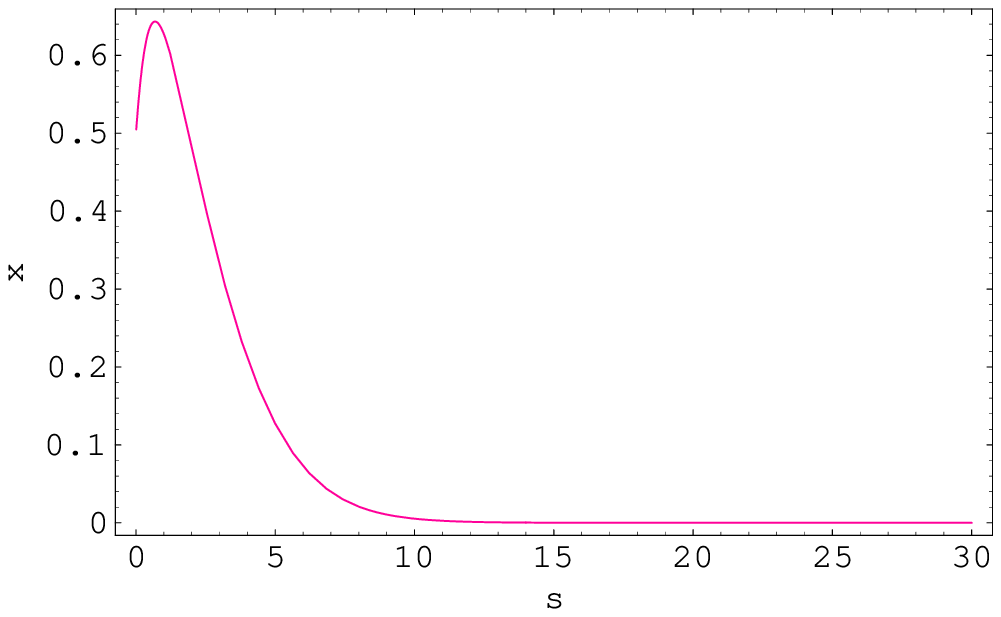,height=2.5in,width=3.2in} \caption{The
evolution of the field $\phi$ with respect to $s$ when the phantom
is dominant} \epsfig{file=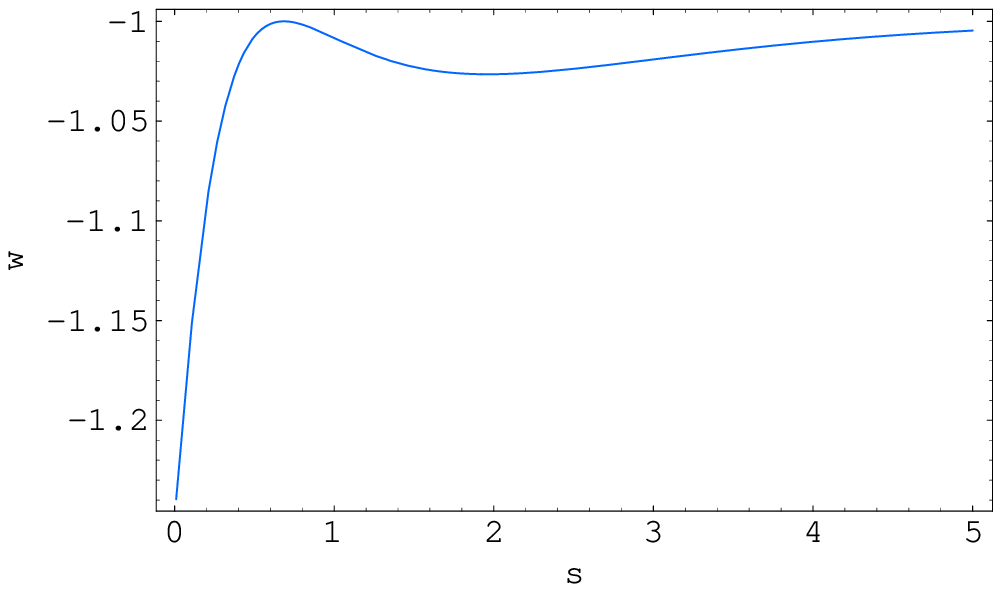,height=2.5in,width=3.2in}
\caption{The evolution of the equation of state of the phantom $w$
with respect to $s$ when the phantom is dominant}
\epsfig{file=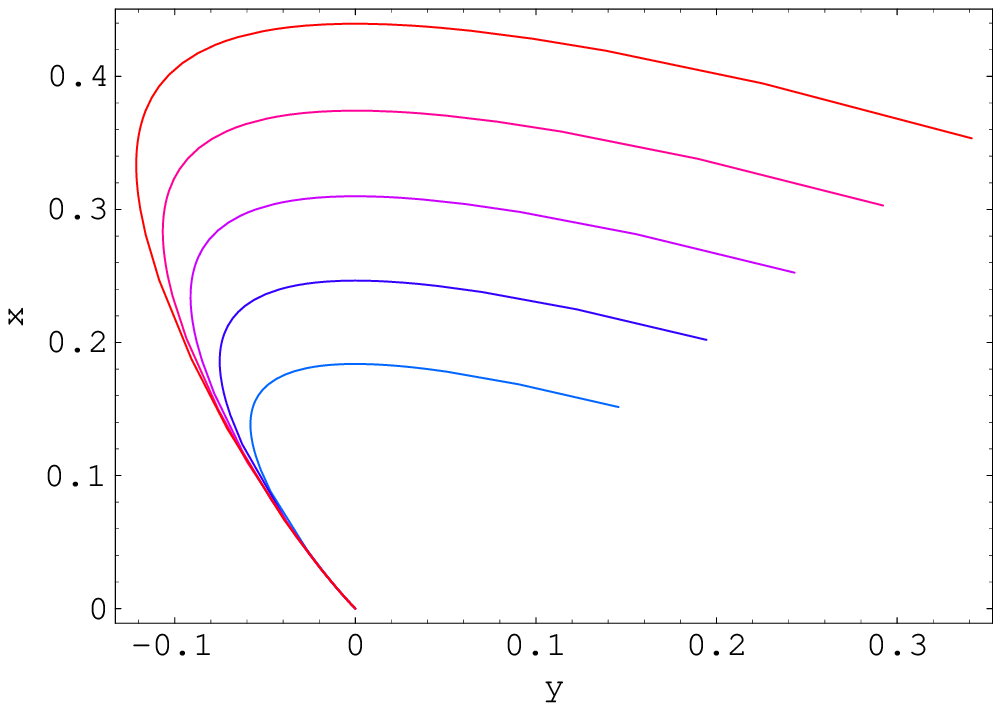,height=2.5in,width=3.2in} \caption{The
attractor property of the system in the phase plane when the
phantom is dominant}
\end{figure}

\vspace{0.4cm} \noindent\textbf{4. Evolution of phantom at the
presence of matter and radiation} \vspace{0.4cm}

In former section, we study the evolution of the phantom when it
dominates over all other energy density. Now, we consider the
radiation and matter density and investigate the evolution of the
phantom field. The presence of radiation and matter will alter the
Eq.(\ref{H}) as
\begin{equation}\label{newH}
H^2=\frac{\kappa}{3}\left(\rho_{\phi}+\rho_M+\rho_r\right)
\end{equation}

\noindent where $\rho_M$ and $\rho_r$ denote the non-relativistic
and relativistic components (or matter and radiation) energy
density respectively. Eq.(\ref{newH}) could be rewritten as
\begin{equation}\label{Hr}
 H^2=H_i^2\left(\frac{\rho_{\phi}}{\rho_{c, i}}+\Omega_{M, i}\left(\frac{a_i}{a}\right)^3
 +\Omega_{r, i}\left(\frac{a_i}{a}\right)^4\right)
\end{equation}

\noindent where the subscript $i$ denotes the quantity at the
initial time $t_i$. $\rho_{c, i}$ is the critical density of the
universe at the initial $i$, which is defined as $\rho_{c,
i}=\frac{3H_{i}^2}{\kappa}$. $\Omega_{M, i}$ and $\Omega_{r, i}$
are the cosmic density parameters for matter and radiation at
$t_i$. We introduce the new dimensionless variables

\begin{eqnarray}\label{newv}
x&&=\frac{\phi}{\phi_0}\nonumber\\y&&=\frac{d\phi}{dt}\nonumber\\N&&=\ln
a
\end{eqnarray}

\noindent Then the Eq.(\ref{T}) will reduce to the following
equation systems
\begin{eqnarray}\label{eqsys}
\frac{dx}{d N}&&=\frac{y}{H_i\phi_0}\xi \nonumber\\
\frac{dy}{d N}&&=-3y(1+y^2)+\frac{(1+y^2)V'}{H_iV}\xi
\end{eqnarray}

\noindent where the prime denote the differentiation with respect
to $x$ and $\xi$ is defined as
\begin{equation}\label{xi}
 \xi=\left[\frac{\kappa V}{3H_i^2\sqrt{1+y^2}}+\Omega_{M, i}\left(\frac{a_i}{a}\right)^3
 +\Omega_{r, i}\left(\frac{a_i}{a}\right)^4\right]^{-1/2}
\end{equation}

\noindent If we specify the initial scale factor $a_i=1$ then we
can express $\xi$ as a function of $N$
\begin{equation}\label{xin}
 \xi(N)=\left[\frac{\kappa V}{3H_i^2\sqrt{1+y^2}}+\Omega_{M,
 i}e^{-3N}
 +\Omega_{r, i}e^{-4N}\right]^{-1/2}
\end{equation}

\noindent Before we carry out the numerical study, we would like
to analyze the system of equations qualitatively. When N goes to
be very large, or at late time, the contribution to $H$ from
matter and radiation will become negligible. Therefore, the
Eq.(\ref{eqsys}) will reduce to

\begin{eqnarray}\label{reducesys}
\frac{dx}{d
N}&&=\sqrt{\frac{3}{\kappa}}\frac{y}{\phi_0}\left[\frac{\sqrt{1+y^2}}{V(x)}\right]^{1/2}\noindent\\
\frac{dx}{d
N}&&=-3y(1+y^2)+\sqrt{\frac{3}{\kappa}}\frac{V'(x)}{V(x)}
\left[\frac{\sqrt{1+y^2}}{V(x)}\right]^{1/2}(1+y^2)
\end{eqnarray}

Note that these equations are re-expressions of the
Eq.(\ref{auto}) in term of $N$ instead of $s$. These two different
expression are essentially the same and therefore the condition
for the existence of attractor solution $V''(x_c)<0$ still hold
true. Thus, from the above qualitative analysis, we can conclude
that for the tachyon potential Eq.(\ref{poten}), there is still a
late time attractor solution even if at the presence of matter and
radiation.

Next, we will numerically study the system at the presence of
matter and radiation in the potential(\ref{poten}) and obtain the
results that will confirm our qualitative analysis. Substitute
Eq.(\ref{poten}) into Eq.(\ref{eqsys}) we have

\begin{eqnarray}\label{fianaleqsys}
\frac{dx}{d N}&&=\frac{y}{H_i\phi_0}\xi'(N)\noindent\\
\frac{dy}{d N}&&=-3y(1+y^2)-\frac{(1+y^2)x}{H_i\phi_0(1+x)}\xi'(N)
\end{eqnarray}

\noindent where $\xi'(N)$ is
\begin{equation}\label{xiprime}
\xi'(N)=\left[\frac{\kappa
V_0(1+x)e^{-x}}{3H_i^2\sqrt{1+y^2}}+\Omega_{M,
 i}e^{-3N}
 +\Omega_{r, i}e^{-4N}\right]^{-1/2}
\end{equation}

Solving these equations will give us some insights into the
evolution of the field and the quantities of interest. We specify
our starting point as the equipartition epoch, at which the
$\Omega_M=\Omega_r=0.5$. The results are shown in Fig.4, Fig.5,
and Fig.6 (The plots are done by choosing the dimensionless
parameter $\frac{\kappa V_0}{3H_i^2}=0.00000033$). From these
results, we can conclude that the current energy density
composition is not an final evolution stage in this specific
model. In stead, the universe will evolve to a de Sitter like
attractor regime in the future and the energy density of dark
energy will become completely dominant over the non-relativistic
matter ($\Omega_{\phi}\rightarrow 1$) and behave as a cosmological
constant.
\begin{figure}
\epsfig{file=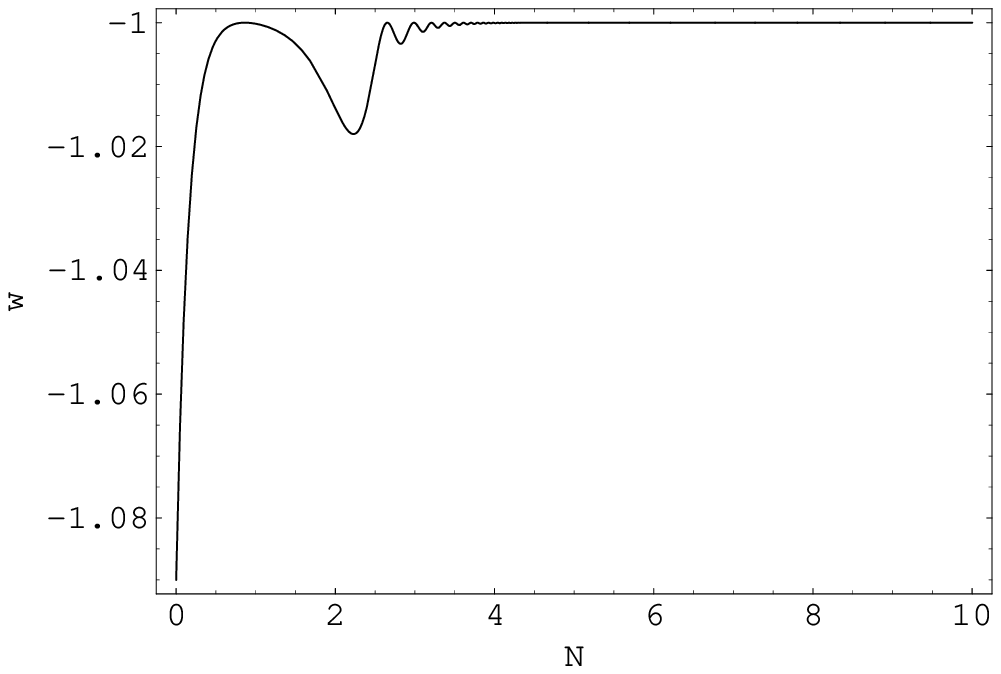,height=2.5in,width=3.2in} \caption{The
evolution of the equation of state of phantom with respect to $N$
at the presence of matter and radiation. Plot begins from the
equipartition epoch}
\epsfig{file=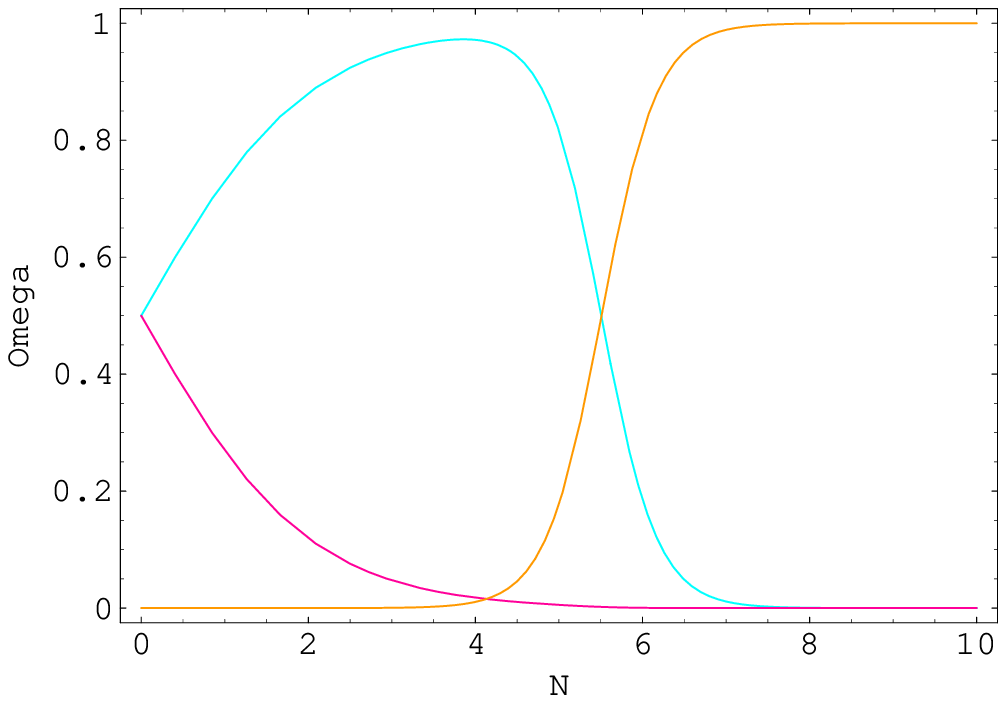,height=2.5in,width=3.2in} \caption{The
evolution of $\Omega_M$(blue curve), $\Omega_r$ (red curve)and
$\Omega_{\phi}$(orange curve) with respect to $N$. The plot begins
from the equipartition epoch ($\Omega_M=\Omega_r$).}
\epsfig{file=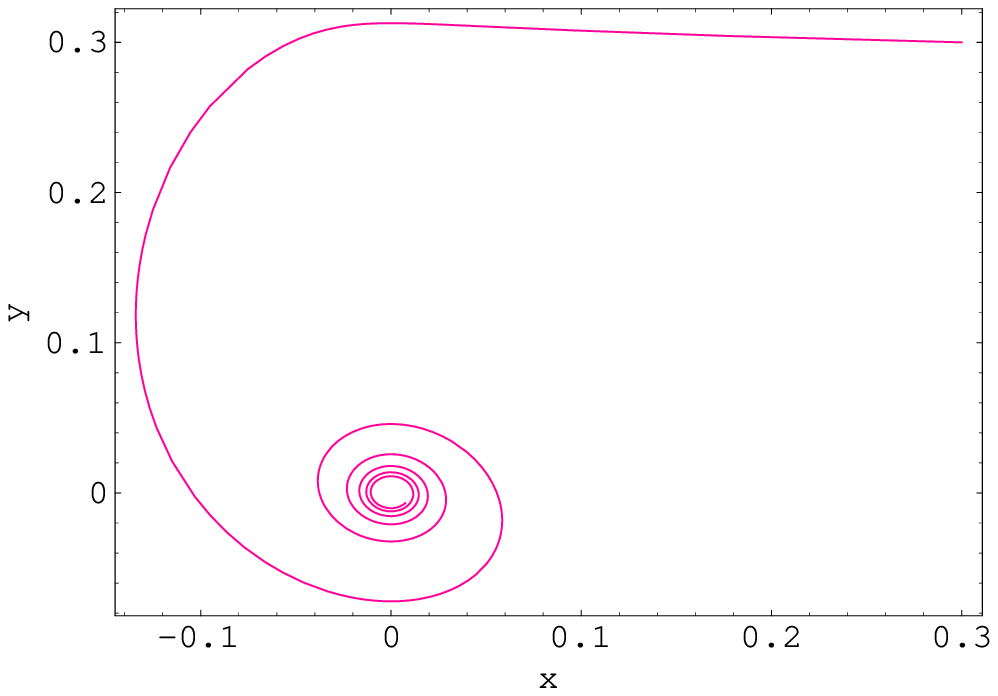,height=2.5in,width=3.2in} \caption{The
attractor property of the phantom at the presence of matter and
radiation. Plot begins from the equipartition epoch.}
\end{figure}

\vspace{0.4cm} \noindent\textbf{5. Discussion} \vspace{0.4cm}

In this paper, we study the cosmological implication of a
Born-Infeld type Lagrangian with negative kinetic energy term. One
may also consider this as a new realization of k-essence with an
equation of state $w\leq-1$. Analysis to the dynamical evolution
of the phantom model indicates that it admits a late time
attractor solution, at which the field behaves as a cosmological
constant. Before the field evolves to its attractor regime, the
equation of state of the field is less than $-1$. In the k-essence
model, the speed of sound is
\begin{eqnarray}\label{speedofsound}
c_s^2=\frac{p_{,X}}{\rho_{,X}}=\frac{L_{\phi,X}}{L_{\phi,X}+2XL_{\phi,X
X}}
\end{eqnarray}

\noindent where $p=L_{\phi}(\phi, X)$ and
$\rho=2XL_{\phi,X}-L_{\phi}(\phi, X)$ with
$X=\frac{1}{2}(\partial_{\mu}\phi)^2$. So, the speed of sound of
the phantom with B-I Lagrangian is $1+\dot{\phi}^2$, which means
that the perturbations of the background field can travel faster
than light as measured in the preferred frame where the background
field is homogeneous. For a time dependent background field, this
is not a Lorentz invariant state. However, it does not violate
causality because the underlying theory is manifestly Lorentz
invariant and it is not possible to transmit information faster
than light along arbitrary space-like directions or create closed
time-like curves\cite{erickson}.

Up to now, the observation data do not tell us what should be the
nature of dark energy. But the future observation will be helpful
to determine whether the dark energy is phantom, quintessence, or
cosmological constant. If the equation of state $w<-1$ is
completely confirmed by observation, then its implication to
fundamental physics would be astounding, since it cannot be
achieved with substance with canonical Lagrangian. Phantom with
B-I Lagrangian could be an interesting candidate for dark energy
with an equation of state $w<-1$.

\vspace{0.8cm} \noindent ACKNOWLEDGEMENT: We thank Alessandro
Melchiorri for helpful comments. This work was partially supported
by National Nature Science Foundation of China under Grant No.
19875016, and Foundation of Shanghai Development for Science and
Technology under Grant No.01JC14035.


\begin{thebibliography}{99}
\bibitem{newobservation} P. de Bernardis et al. Nature {\bf 404} 955(2000); S. Hanany et al.
Astrophys. J. {\bf 545} 1 (2000); N. Bahcall, J. P. Ostriker, S.
Perlmutter and P. J. Steinhardt Science {\bf 284} 1481(1999); S.
Perlmutter et al. Astrophys. J. {\bf 517} 565(1999 ); A. G. Riess
et al., Astron. J. {\bf 116} 1009(1998)
\bibitem {steinhardt} B. Ratra and P. J. Peebles, Phys. Rev. {\bf D37} 3406(1988);
R. R. Caldwell, R. Dave and P. J. Steinhardt, Phys. Rev. Lett.
{\bf 80} 1582(1998); P. J. Steinhardt, L . Wang and I . Zlatev,
Phys. Rev. {\bf D59} 123504(1999); I. Zlatev, L. Wang and P. J.
Steinhardt, Phys. Rev. Lett. {\bf 82} 896(1999); K. Coble, S.
Dodelson, J. Frieman, Phys. Rev. {\bf D55} 1851(1997); X. Z. Li,
J. G. Hao, D. J. Liu, Class.Quant.Grav. 19
6049(2002)
\bibitem{melchiorri} A. Melchiorri, L. Mersini, C. J.
Odmann and M. Trodden, Astro-ph/0211522
\bibitem{tonry} J. L. Tonry \textit{et al.}, astro-ph/0305008
\bibitem {caldwell1} R.R. Caldwell, Phys.Lett. \textbf{B545} 23(2002).
\bibitem {sahni} V. Sahni and A. A. Starobinsky, Int. J. Mod. Phys. \textbf{D9} 373(2002)
\bibitem {parker} L. Parker and A. Raval, Phys. Rev. \textbf{D60} 063512(1999)
\bibitem {chiba} T. Chiba, T. Okabe and M. Yamaguchi, Phys. Rev. \textbf{D62} 023511(2000)
\bibitem {boisseau} B. Boisseau, G. Esposito-Farese, D. Polarski and A. A. Starobinsky, Phys. Rev. Lett.\textbf{85} 2236, (2000)
\bibitem {schulz} A. E. Schulz, Martin White, Phys.Rev. \textbf{D64} 043514(2001)
\bibitem {faraoni} V. Faraoni, Int. J. Mod. Phys. \textbf{D64} 043514 (2002)
\bibitem {maor} I. Maor, R. Brustein, J. Mcmahon and P. J. Steinhardt, Phys. Rev. \textbf{D65} 123003(2002)
\bibitem {onemli} V. K. Onemli and R. P. Woodard, Class. Quant. Grav. \textbf{19} 4607(2002)
\bibitem {torres} D. F. Torres, Phys. Rev. \textbf{D66} 043522 (2002)
\bibitem {carroll} S. M. Carroll, M. Hoffman, M. Trodden, astro-ph/0301273
\bibitem {frampton} P. H. Frampton, Stability Issues for $w < -1$ Dark Energy, hep-th/0302007
\bibitem {hao} J. G. Hao and X. Z. Li, gr-qc/0302100, to be published in Phys. Rev. \textbf{D};
X. Z. Li and J. G. Hao, hep-th/0303093.
\bibitem {caldwell2} R. R Caldwell, M. Kamionkowski and N. N. Weinberg, astro-ph/0302506
\bibitem {gibbons}  G. W. Gibbons,  hep-th/0302199
\bibitem {feinstein}A. Feinstein and S. Jhingan,
 hep-th/0304069; L. P. Chimento and A. Feinstein,
astro-ph/0305007; P. Singh, M. Sami and N. Dadhich, hep-th/0305110
\bibitem {tachyon}A. Sen, \textit{JHEP} \textbf{0204}, 048 (2002); A. Sen,
\textit{JHEP} \textbf{0207}, 065(2002); G.W.Gibbons, \textit{Phys.
Lett.} \textbf{B537}, 1 (2002); M. Fairbairn and M. H. G. Tytgat,
Phys.Lett.\textbf{B546}, 1 (2002); A. Sen \textit{J. Math. Phys.}
\textbf{42}, 2844(2001); G. W. Gibbons, K. Hori  and P. Yi,
\textit{Nucl. Phys.} \textbf{B596}, 136 (2001); A. Sen,
hep-th/0303057; J. G. Hao and X. Z. Li, \textit{ Phys. Rev.}
\textbf{D66}, 087301(2002); X. Z. Li and X. H. Zhai, \textit{Phys.
Rev.}\textbf{D67}, 067501(2003); A. Frolov, L. Kofman and A.
Starobinsky, hep-th/0204187; S. Mukohyama, \textit{ Phys. Rev.}
\textbf{D66}, 024009(2002); T. Padmanabhan, \textit{Phys. Rev.}
\textbf{D66}, 021301 (2002); M. Sami and T. Padmanabhan,
\textit{Phys. Rev.} \textbf{D67}, 083509(2003); G. Shiu and I.
Wasserman, \textit{Phys. Lett.} \textbf{B541}, 6(2002); L. Kofman
and A. Linde, hep-th/020512; H. B. Benaoum, hep-th/0205140; A.
Ishida and S. Uehara, hep-th/0206102; T. Chiba, astro-ph/0206298;
T, Mehen and B. Wecht, hep-th/0206212; A. Sen, hep-th/0207105; N.
Moeller and B. Zwiebach,\textit{JHEP} \textbf{0210}, 034(2002); J.
M. Cline, H. Firouzjahi and P. Martineau, hep-th/0207156; S.
Mukohyama, hep-th/0208094; P. Mukhopadhyay and A. Sen,
hep-th/020814; T. Okuda and S. Sugimoto, hep-th/0208196; G.
Gibbons , K. Hashimoto and P. Yi, hep-th/0209034; M. R. Garousi,
hep-th/0209068; A. Sen, hep-th/0209122; B. Chen , M. Li and F.
Lin, hep-th/0209222; J. Luson, hep-th/0209255; C. Kim, H. B. Kim,
Y. Kim and O. K. Kwon, hep-th/0301142; X.Z. Li, D.J.Liu and
J.G.Hao, hep-th/0207146; J.M.Cline, H. Firouzjahi and P.
Martineau, hep-th/0207156; G. Felder, L. Kofman and A.
Starobinsky, \textit{JHEP} \textbf{0209}, 026(2002); S. Mukohyama,
arXiv:hep-th/0208094; G.A. Diamandis, B.C. Georgalas , N.E.
Mavromatos, E. Papantonopoulos, hep-th/0203241; G.A. Diamandis,
B.C. Georgalas , N.E. Mavromatos, E. Papantonopoulos, I. Pappa,
hep-th/0107124; M. C. Bento, O. Bertolami and A. A.
Sen,hep-th/020812;  M.C. Bento, O. Bertolami., A. A. Sen,
\textit{Phys.Rev.} \textbf{D67}, 023504,2003;  C. Kim , H. B. Kim
and Y. Kim, hep-th/0210101;  C. Kim, Y. Kim, O. K. Kwon, C. Oh
Lee, hep-th/0305092 ; H. Lee, W. S. l'Yi, hep-th/0210221;
J.S.Bagla,  H.K.Jassal, T.Padmanabhan, astro-ph/0212198; M. Sami,
P Chingangbam and T Qureshi, hep-th/0301140; M. Sami, P
Chingangbam and T Qureshi, Phys.Rev. \textbf{D66} 043530(2002); M.
Sami, Mod.Phys.Lett. \textbf{A18} 691(2003); F. Leblond and  A. W.
Peet, hep-th/0303035; F. Leblond, A. W. Peet, hep-th/0305059; T.
Matsuda, hep-ph/0302035; T. Matsuda, hep-ph/0302078; A. Das and A.
DeBenedictis, gr-qc/0304017; M. Majumdar, A. Davis,hep-th/0304226;
X. Z. Li, J. G. Hao and D. J. Liu, \textit{Chin.Phys.Lett.}
\textbf{19}, 1584(2002); G.W. Gibbons, hep-th/0301117
\bibitem {li}X. Z. Li, J. G. Hao and D. J. Liu,
\textit{Chin.Phys.Lett.} \textbf{19},(2002)1584;
\bibitem {gibbons1}  G.W. Gibbons, Thoughts on
tachyon cosmology, hep-th/0301117
\bibitem{kutasov} D. Kutasov, M. Marino and G. W. Moore,
\textit{JHEP} \textbf{045}, 0010(2000)
\bibitem {erickson} J. K. Erickson et al, Phys.Rev.Lett. \textbf{88},121301(2002)
\end{thebibliography}
\end{document}